\documentclass[aps,twocolumn,showpacs]{revtex4}
\usepackage{epsfig}

\begin{document}
\title{
Temperature-induced pair correlations in clusters and nuclei}
\author{S. Frauendorf$^{1,2}$, N.K. Kuzmenko$^3$,
V.M. Mikhajlov$^4$,
and J.A. Sheikh$^2$\\
}

\affiliation{$^1$Department of Physics, University of Notre Dame,
Notre Dame, IN 46556, USA}
\affiliation{$^2$IKH Forschungszentrum Rossendorf,
 PF 510119, 01324 Dresden, Germany}
\affiliation{$^3$V.G. Khlopin Radium Institute, 194021,
St.-Petersburg, Russia}
\affiliation{$^4$Institute of Physics St.--Petersburg State
University 198904, Russia}

\date{\today}

\begin{abstract}
The pair correlations in mesoscopic systems such as $nm$-size
superconducting
clusters and nuclei are studied at finite temperature for the
canonical ensemble of fermions in model spaces with a fixed particle number:
i) a degenerate spherical shell (strong coupling
limit), ii) an equidistantly spaced deformed shell
(weak coupling limit).
It is shown that
after the destruction of the pair correlations  at $T=0$ by a strong
magnetic field or rapid rotation,  heating can bring them
 back. This phenomenon is a consequence of the fixed
number of fermions in the canonical ensemble.
\end{abstract}


\maketitle

The pair correlations in a macroscopic superconductor are
destroyed by increasing the temperature or the external magnetic
field. The critical field which marks the boundary between the
superconducting and normal phases, is a decreasing function of
the temperature $T$. The BCS theory, which is the mean field
approximation based on the grand canonical ensemble, describes
very accurately this regime. Applying the grand canonical mean
field approach to rotating nuclei \cite{egido}, where the angular
velocity plays the role of the magnetic field, gives an analogous
result: The angular velocity where the pair correlations
disappear decreases with the temperature. Nuclei and atomic
nano-size clusters are composed of a small and fixed number of
particles, the single-particle spectrum is discrete  and the
level spacing is comparable with the pair gap. Due to these
facts, the fluctuations of the order parameter become important,
which smear out the transition from superconducting to normal
phase \cite{egido,perenboom,delftralph,moretto,shim} and lead to
pronounced differences between a system with even and odd
particle number \cite{balian}.

In order to properly take into account these fluctuations
one has to use the canonical ensemble. The most direct way is to
calculate the partition function from the exact eigenvalues of
the Hamiltonian, which is possible for some models. In this
Letter we study the  simple model of fermions occupying an
isolated shell of single particle states and interacting by a
pairing force. We shall demonstrate that at zero temperature the
magnetic field or the angular velocity attenuate the paring in a
step-wise manner until it disappears completely above a critical
value. {\it For a mesoscopic system in the strong magnetic field
heating may bring back the pair correlations.} This surprising
effect is a consequence of the fixed number of fermions in such
a small system. The reduction of the fluctuations in particle number
leads to a strong increase of the  fluctuations of the order parameter,
which constitute the pair correlations above the critical field.
A re-entrance of pair correlations has first been discussed
by Balian, Flocard and Veneroni \cite{balian}, who 
studied ensembles with either only even numbers of particles  or
only odd numbers of particles. We shall demonstrate that the
more stringent restriction to a  fixed number
of particles, which is realized in small
systems, leads to qualitatively different results.

First we consider fermions in a spherical potential.
Then the  electronic states in the cluster (the spin-orbit coupling
can be neglected) are characterized by $l$, the orbital momentum of the shell,
its $z$-projection $\lambda$ and spin projection $\sigma$, which we
label by
$k=\{\lambda ,\sigma\}$. The condition, $k>0$, used below in
Eq.~({\ref{ham}}),
means that $\lambda + \sigma>0$. Due to the  strong spin-orbit splitting in
nuclei the nucleonic states in a shell are specified
by the total angular momentum $j$,
($\vec j=\vec l + \vec s$), and its projection $m\equiv k$.
We assume that there are only pair correlations within the last partially
filled shell. The corresponding "single-shell" Hamiltonian
\begin{eqnarray}\label{ham}
H=H_{\mathrm{pair}}-\omega M,~~~M= L_z+gS_z, \\
H_{\mathrm{pair}}=-GA^+A,~~~
A^+=\sum_{k>0} a^+_ka^+_{\bar {k}}, \nonumber
\end{eqnarray}
consists of the pairing interaction $H_{\mathrm{pair}}$, which
acts between the valence fermions in the last shell with the
strength $G$, and the ``cranking term''. The $z$-components of
the total orbital angular momentum and spin, which are the sums
of all valence fermion contributions, are denoted by $L_z$ and
$S_z$. The operator $A^+$  creates a fermion pair in the
time-reversal states $(k,\bar {k})$. In the case of clusters, the
cranking term represents the interaction  of the electrons  with
the magnetic-field, where we introduce the Larmour frequency
$\omega=\mu_BB$ and the Bohr magneton $\mu_B$. The gyromagnetic
ratio $g=2$. In the case of nuclei, the cranking term takes into
account the rotational perturbations at the angular frequency
$\omega$ and $g=1$.

The eigenvalues of $H_{\mathrm{pair}}(\omega=0)$
are  \cite{esebag}
\begin{equation}\label{Enu}
E_{\nu}=-\frac{G}{4}(N_{\mathrm{sh}}-\nu)(\Omega+2-
N_{\mathrm{sh}}-\nu).
\end{equation}
were $N_{\mathrm{sh}}$ is the number of  particles in the shell,
which consists of $\Omega$ degenerated single-particle states and
$\Omega=4l+2$ for clusters and $2j+1$ for nuclei.
The seniority $\nu$, which is the number of
unpaired particles, is constrained by
$0 \le \nu \le N_{\mathrm{sh}}$.
We assume that $\nu\leq \Omega/2$, otherwise one can pass to the
hole representation. Each $\nu$-state at $\nu\ge 1$ is
degenerated. The eigenvalues of the Hamiltonian  (\ref{ham})
are
\begin{equation}\label{Unu}
E_{\nu,i}(\omega)=E_{\nu} - \omega M_{\nu,i}.
\end{equation}
The index $i$ involves additional quantum numbers of the $\nu$-state
including the spin and possible orbital momenta and their
$z$-projections for clusters or the total angular momentum of
$\nu$ nucleons and its projection for nuclei. In the case of clusters,
$\mu_B M_{\nu ,i}$ is the total magnetic moment of $\nu$ particles
while in nuclei $M_{\nu ,i}$ is the angular momentum $z$ projection
of $\nu$ nucleons. For
simplicity, we refer to $M$ as the angular momentum and omit
the index $i$ for the states with minimal energy at each $\nu$, which have
maximal $M_{\nu}$.

The canonical partition function $Z$ and the mean value $\bar M$
are given by
\begin{eqnarray}\label{Z}
Z(T,\omega)=\sum_{\nu,i}\exp(-\frac{ E_{\nu,i}(\omega)}{T}),\\
\bar M(T,\omega)=\frac{T}{Z}\frac{\partial Z}{\partial\omega}.
\end{eqnarray}                
The evaluation uses
the fact that for each $\nu$ the sum over $i$  can be reduced to a
sum over single-particle projections.
For  nuclei,
the numerical diagonalization procedure described in Ref. \cite{sheikh}
is used, which permits us also to treat a non-degenerate shell
(see below).
 We consider only the part of $\bar M$ which is generated by
the particles near the Fermi surface that participate in the
pair correlation.  The contributions for the other particles can
be found in ref. \cite{kuzmenko2} and will be discussed together
with the details of the evaluation of the sums for clusters in an
forthcoming extended paper \cite{kuzmenko1}.

We introduce the ``canonical'' pair gap $\Delta_{\mathrm{can}}$
as a measure of the correlation energy
\begin{eqnarray}\label{dcan}
\Delta_{\mathrm{can}}^2(T,\omega)
=\frac{G}{Z(G)}\sum_{\nu,i}\langle \nu,i|A^+A|\nu,i\rangle
\exp(-\frac{ E_{\nu,i}(G)}{T})
\nonumber \\
-\frac{G}{Z(G=0)}\sum_{\nu,i}\langle\nu,i,0|A^+A|\nu,i,0
\rangle\exp(-\frac{ E_{\nu,i}(G=0)}{T}),
\end{eqnarray}
where $|\nu,i,0\rangle$ denote the uncorrelated fermion
configurations in the shell. The second term subtracts the
expectation value of the pairing interaction in an ensemble of
uncorrelated fermions. A detailed discussion of the proper
definition of $\Delta_{\mathrm{can}}$ was given in the review
\cite{delftralph}. Applying the mean-field approximation and the
grand canonical ensemble to our model, the thus introduced
$\Delta_{\mathrm{can}}$  becomes the familiar BCS gap parameter
$\Delta_{\mathrm{mf}}$. However, $\Delta_{\mathrm{can}}$ must be
clearly distinguished from $\Delta_{\mathrm{mf}}$ because it
incorporates the correlations caused by the fluctuations of the
order parameter.
\begin{figure}[t]
\mbox{\psfig{file=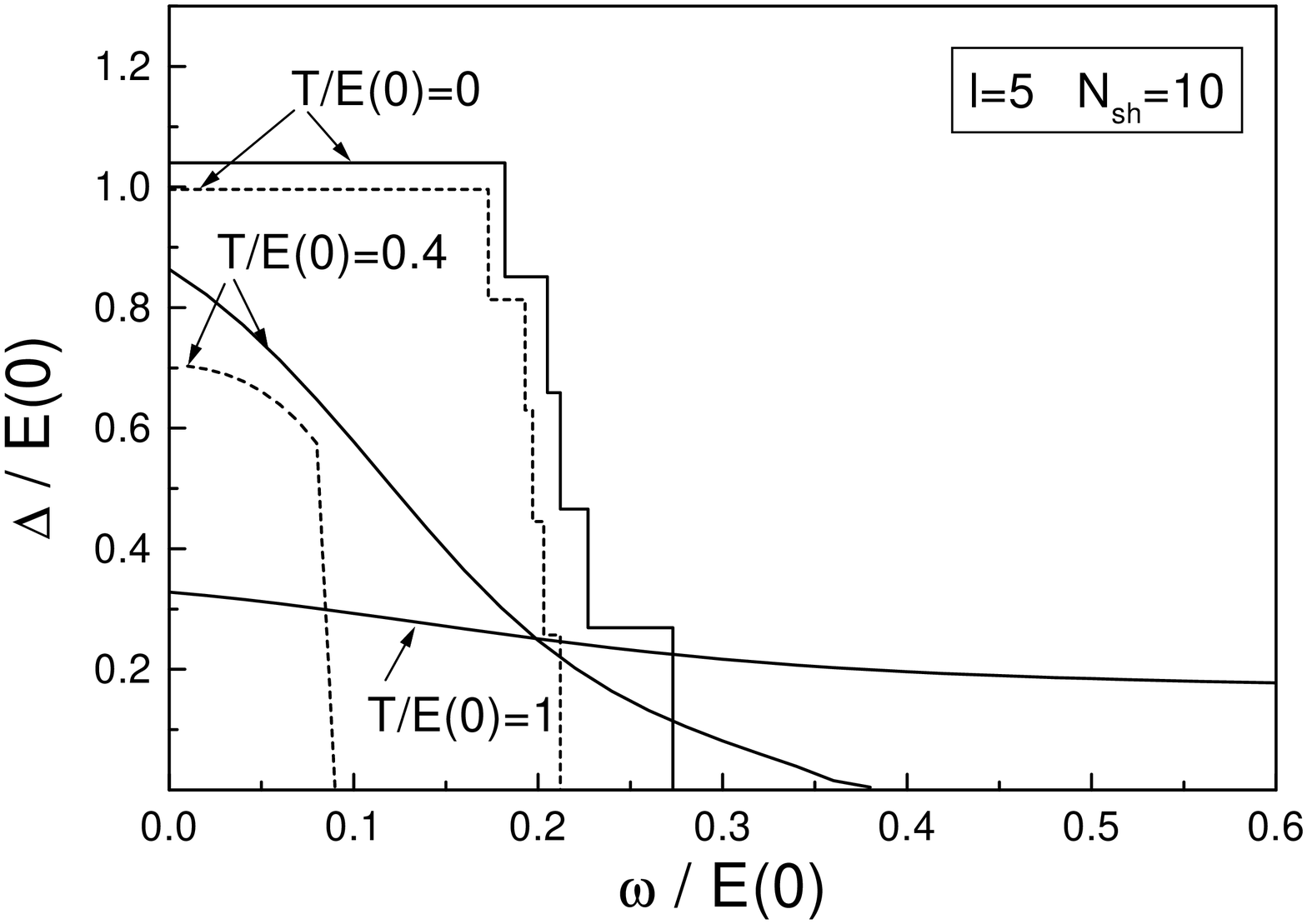,width=8cm}}
\caption
{ {\label{f:delomcl}}
 Canonical gap  $\Delta_{\mathrm{can}}(T,\omega)$
(full lines) and the mean-field
gap $\Delta_{\mathrm{mf}}(T,\omega)$ (dotted lines)
v.s. the frequency $\omega$ for a spherical shell.}
\end{figure}
We take as energy scale $E(0)=G\Omega/4$,
the quasiparticle energy at $T=\omega=0$.
A value of $E(0)=0.3-0.4~meV$ was found for
Al-clusters with radii $R=5-10~nm$ \cite{ralph}, which sets the energy scale.
The nuclear mass measurements give $E(0)=0.8-1.5~MeV$.
\begin{figure}
\noindent
\mbox{\psfig{file=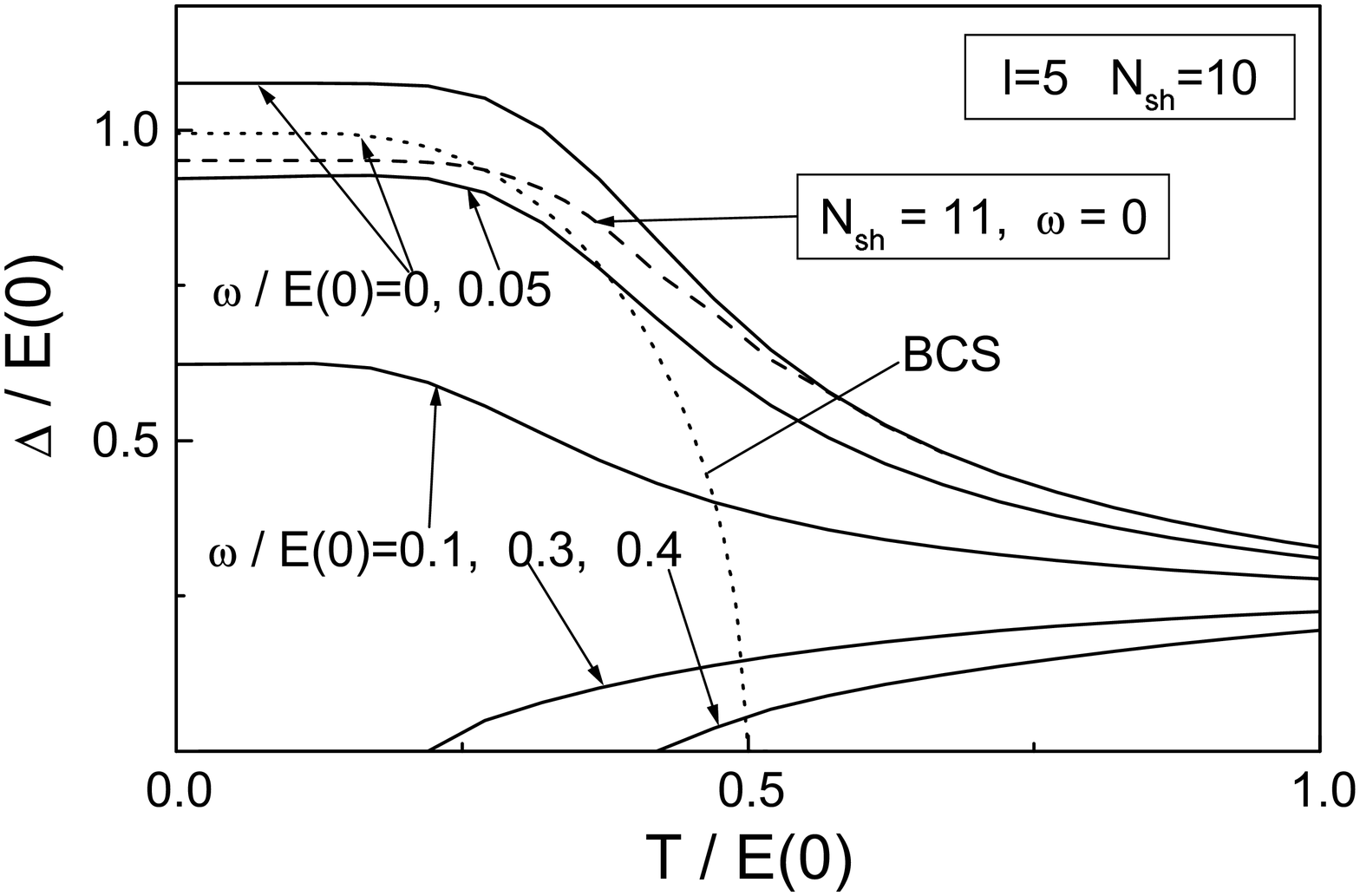,width=8cm}}
\caption
{ {\label{f:deltsph}}
 Canonical gap  $\Delta_{\mathrm{can}}(T,\omega)$
for even (full lines) and odd  (the dashed line) particle number,
and the mean-field gap
$\Delta_{\mathrm{mf}}(T,\omega)$ (dotted line -BCS)
v.s. the temperature $T$ for a spherical shell. }
\end{figure}
Let us first consider the destruction of the pair correlations at
$T=0$. Since $M_{\nu}$ increases with $\nu$,
the energy $E_{\nu}(\omega)$ becomes smaller than
$E_{\nu -2}(\omega)$ at the frequency
\begin{equation}\label{omenu}
\omega_\nu ( M_{\nu} - M_{\nu-2} ) = E_{\nu}-E_{\nu -2}.
\end{equation}
When the state of the lowest energy changes
from $\nu - 2$ to $\nu$, two more electron states are blocked and
the pair correlations are reduced.
The last step at the critical frequency
$\omega_{\mathrm{crit}}$ leads to the maximal seniority
$\nu_{\mathrm{max}}$, where
all particle states are blocked and the pairing is completely
destroyed.
Fig.~\ref{f:delomcl} illustrates
 the step-wise destruction of pairing by
blocking for a half-filled  $l=5$ shell, which is  the Fermi level
in a spherical Al-cluster with about $10^3$ atoms.

Fig.~\ref{f:delomcl}  also shows
the gap $\Delta_{\mathrm{mf}}(\omega,T)$ obtained by applying
the mean-field  approximation
(cf.~\cite{esebag,kuzmenko2}) to the single-shell model. For $T=0$,
the pair correlations are more rapidly destroyed than
for the exact solution. The
quantal fluctuations of the order parameter generate additional
pair correlations. For finite $T$, the mean-field gap
 behaves as known from macroscopic
superconductors:  $\Delta_{\mathrm{mf}}(\omega,T)$ is reduced at
$T=0.4 E(0)$ and reaches $0$ at a lower value of $\omega$. This is
the familiar shift of the critical frequency
$\omega_{\mathrm{crit}}$ toward smaller values  with increasing
$T$. However, the canonical gap $\Delta_{\mathrm{can}}(\omega,T)$
behaves differently: The abrupt drop around
$\omega_{\mathrm{crit}}$ is smoothed out by the fluctuations of
the order parameter. Moreover { \em substantial pair correlations
appear in the region above $\omega_{\mathrm{crit}}$, which
increase with $T$}. The comparison with
$\Delta_{\mathrm{mf}}(\omega,T)$ shows that the fluctuations
contribute more at finite $T$ than at $T=0$. For $T=E(0)$,  the
pair correlations fall off very gradually with $\omega$. The
nuclear case is quite similar.

Fig.~\ref{f:deltsph} shows how these
temperature-induced pair correlations appear with increasing
$T$. For $\omega=0$, there is a pronounced drop  of
$\Delta_{\mathrm{can}}$ around
$T_{\mathrm{crit}}(\omega=0)=E(0)/2$, where the mean-field  gap
$\Delta_{\mathrm{mf}}$ goes to zero.
Above this
temperature there is a long tail of  pair correlations caused by the
fluctuations. For
$\omega=0.05~E(0)$, the drop is shifted to smaller $T$ by about
the same amount as the $T_{\mathrm{crit}}$ of the mean-field
solution (not shown).
This trend continues for $\omega=0.1~E(0)$.
For larger $\omega\geq \omega_{\mathrm{crit}}$, {\em the
pair-correlations  built up with increasing $T$}.

The temperature-induced pairing can be understood in the following way.
At $T=0$, all electrons are unpaired when the
state of the maximum seniority becomes the ground state for
$\omega>\omega_{\mathrm{crit}}$. At $T>0$, excited states with lower
seniorities enter
the canonical ensemble, which reintroduce the pair correlations.

The degenerate spherical shell corresponds to the strong coupling
limit ($\Delta/d\gg 1$, $d$ distance between the levels) of
ref.~\cite{balian}. For $\omega=0$, one may compare
$\Delta_{\mathrm{can}}$ of the present work with $\Delta_F$
obtained for the ensemble with good particle number parity in
ref.~\cite{balian}, which we will refer to as BVF in what
follows. For $T<0.5~T_{\mathrm{crit}}$, both are similar. However
for $T>T_{\mathrm{crit}}$, $\Delta_F=0$ whereas
$\Delta_{\mathrm{can}}$ remains finite up much higher
temperatures. The case $\omega\not=0$ is not shown in
ref.~\cite{balian} for the strong coupling limit.

In order to investigate the consequence of deviations from the
spherical symmetry, we used the deformed shell-model described
in \cite{sheikh}.
The Hamiltonian is given by
\begin{equation}\label{hamdef}
H=\sum_k e_ka^+_ka_{k}+H_{\mathrm{pair}}-\omega M.
\end{equation}
The solutions are found by numerical diagonalization in
the configuration space of a $j$-shell.

\begin{figure}[t]
\noindent
\mbox{\psfig{file=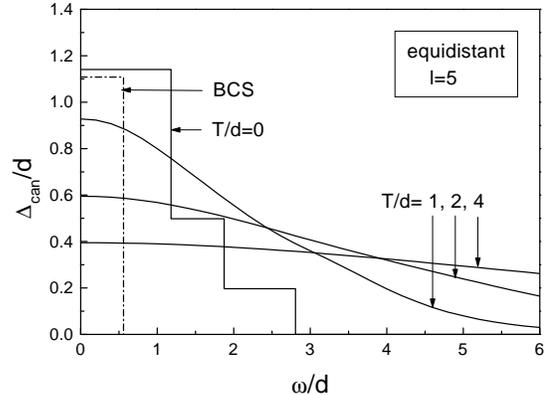,width=8cm}}
\caption{ {\label{f:eqcl}}
 Canonical gap  $\Delta_{\mathrm{can}}(T,\omega)$ for an equidistantly
spaced  $l=5$ shell in a cluster.  The mean-field
gap  $\Delta_{\mathrm{mf}}(T,\omega=0)$ is shown by the dash-dotted line (BCS).
 }
\end{figure}

\begin{figure}[t]
\mbox{\psfig{file=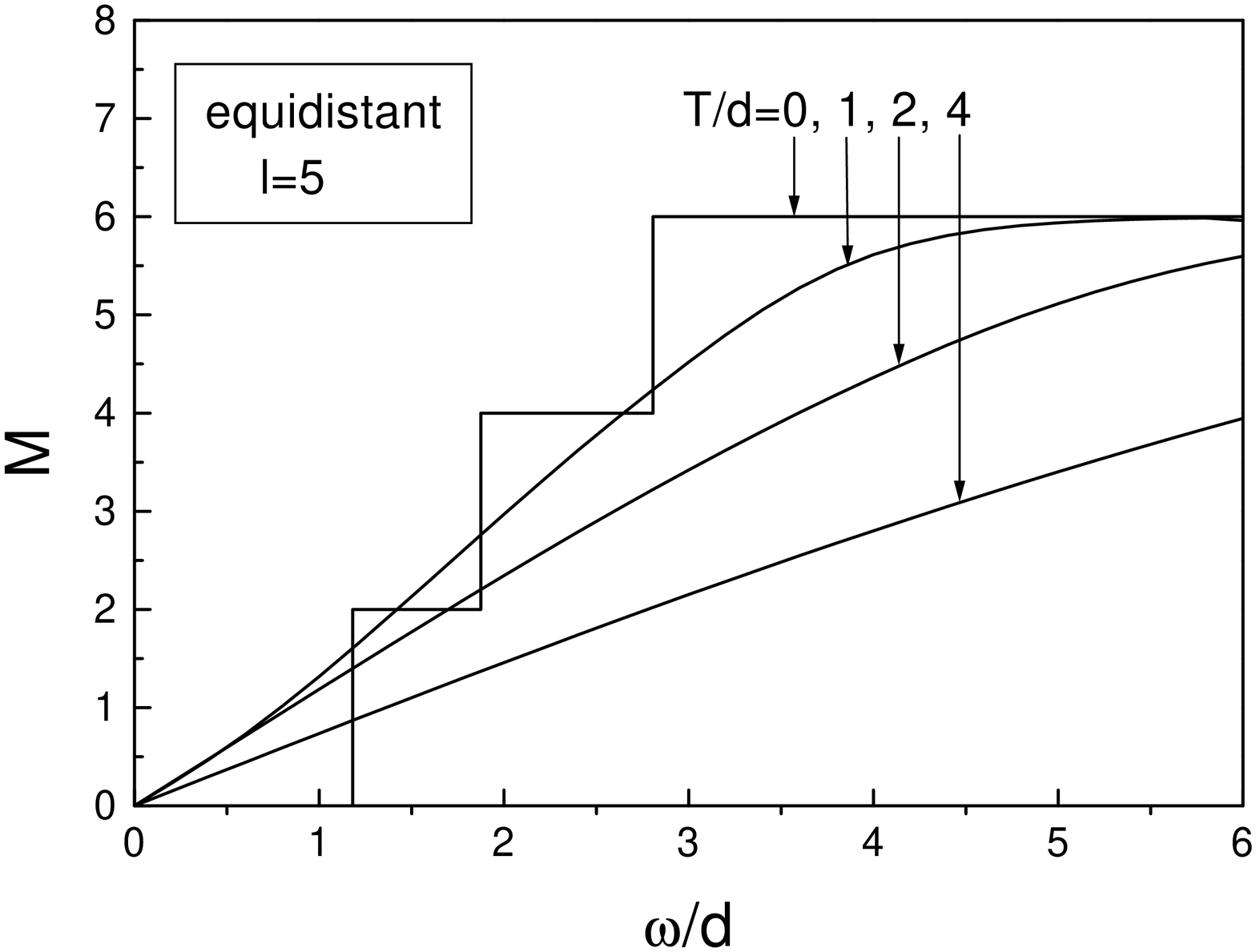,width=8cm}}
\caption{ {\label{f:momcl}}
Angular-momentum $\bar M(T,\omega)$ for an equidistantly
spaced  $l=5$ shell in a cluster.}
\end{figure}

\begin{figure}
\mbox{\psfig{file=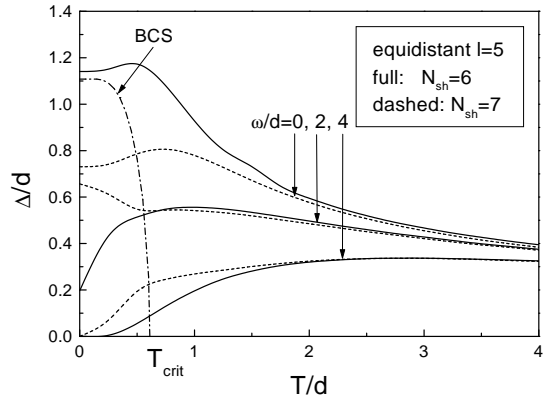,width=8cm}}
\caption{ {\label{f:delnuceo}}
Canonical gap $\Delta_{\mathrm{can}}(T,\omega)$ for an equidistantly
spaced  $l=5$ shell in a cluster with even (full line) and
odd (dotted line) particle number. The mean-field
gap  $\Delta_{\mathrm{mf}}(T,\omega=0)$ is shown by the dash-dotted line (BCS).
}
\end{figure}

Figs.~\ref{f:eqcl} and \ref{f:momcl}  illustrate the case of a
cluster without spherical symmetry. The half- filled  shell
consists of 12 equidistant levels, each of which contains two
states with spin up and down. Our choice of the level distance
$d=0.85~E(0)$ corresponds to weak coupling $\Delta\approx d$ and
is realistic for the nano-clusters. We assume an irregular
cluster shape. As a consequence, the orbital momentum is quenched
and only the spins contribute to the magnetic moment $\mu_B M$.
The behavior of $\Delta_{\mathrm{can}}$ and $\bar M(T,\omega)$ is
similar to the spherical case. The step-wise destruction  of the
pair correlations corresponds to subsequent spin flips of
electrons, which increase $M$ by 2. The temperature-induced
pairing appears in the region above  $\omega\sim 2~d$. The spin
flips have been observed in tunneling experiments on $Al$
nano-clusters \cite{delftralph}. The mean-field gap
$\Delta_{\mathrm{mf}}(T=0)$ breaks down at the first flip, as
discussed by Braun et al. \cite{braun} in their analysis of the
tunneling spectra based on the mean-field approximation. However,
$\Delta_{\mathrm{can}}(T=0)$ is more stable and disappears only
after all levels are blocked by subsequent spin flips.
Fig. \ref{f:delnuceo} demonstrates that the temperature
dependence of $\Delta_{\mathrm{can}}$ is qualitatively similar to
the strong coupling case in Fig. \ref{f:deltsph}. A new feature is
that sometimes $\Delta^{\mathrm{odd}}_{\mathrm{can}}>
\Delta^{\mathrm{even}}_{\mathrm{can}}$. For the curves
$\omega=2d$ in Fig.~\ref{f:delnuceo}, this happens because the
frequency is above the first crossing in the even system (cf.
Fig. \ref{f:eqcl}) but still below the first crossing in the odd
system, which means two states are blocked in the even but only
one in the odd system. For $\omega=4d$ the analogous happens at
the third crossing. In the region of high temperatures
($T>2T_{\mathrm{crit}}$) the canonical pairing gaps for odd and
even systems practically coincide because all members
of $Z$ contribute with a similar weight. At small $T$
they differ because only
the states with small seniorities are  important.

Our choice of parameters lies between to the weak coupling cases
$\Delta/d=1.14$ and 1.19 of the grand canonical
ensembles with even or odd particle number
 studied by  BVF ($\Delta/d= w_F\Delta$ in \cite{balian}).
As in the case of strong coupling, $\Delta_{\mathrm{can}}$
decreases very gradually for $T>T_{\mathrm{crit}}$, whereas
$\Delta_F$ of BVF drops sharply to zero at $T_{\mathrm{crit}}$.
However, also in the region $T<T_{\mathrm{crit}}$
 the results of the  two approaches are different.
The differences between the even and odd systems are much less
pronounced in the canonical ensemble. For $\omega=0$,
$\Delta/d=1.14$, and odd particle number, BVF find $\Delta_F=0$
for $T<0.2~T_{\mathrm{crit}}$, $\Delta_F\not=0$ for
$0.2~T_{\mathrm{crit}}<T<T_{\mathrm{crit}}$, and $\Delta_F=0$ for
$T>T_{\mathrm{crit}}$, which they called
 the re-entrance phenomenon of pairing.
 As seen in Fig. \ref{f:delnuceo}, the strong fluctuations
in the canonical ensemble keep $\Delta_{\mathrm{can}}$  finite in
the whole temperature interval, i. e. there is no re-entrance of
pairing for $\omega=0$ and odd particle number.  On the other
hand, the canonical ensemble gives temperature-induced pair
correlations for $\omega>2~d$ both for  even and odd particle
number, which has not been found for grand-canonical ensembles 
with fixed the particle number parity.
 It should be mentioned that BVF considered a systems with
$N=100,~101$ whereas we studied systems with $N\sim~10$, for which
the conservation of particle number is more important.

We have also studied the case of a half-filled $j=11/2$ shell in
a deformed  axial nucleus by assuming  that $e_k$ in Eq.~(\ref{hamdef})
is proportional to $k^2$.
 We chose
the distance $e_{7/2}-e_{5/2}=0.28~E(0)$, which is realistic. If
the axis of rotation is parallel to the symmetry axis, the
behavior is similar to the deformed clusters, except the steps in
$M$ are different from 2. If the axis of rotation is
perpendicular to the symmetry axis, the projection $M$ of the
angular momentum  is no longer conserved. Then $\bar
M(T=0,\omega)$ is no longer a step function and
$\Delta_{\mathrm{can}}(\omega,T=0)$ decreases in a gradual manner.
The  increase of $\Delta_{\mathrm{can}}(\omega,T)$  with $T$
is found to be weak.
  Hence, temperature-induced pairing is expected
in nuclei that build up large angular-momentum by aligning the
individual angular-momenta of the nucleons near the Fermi surface.
These are either spherical nuclei (see
Figs.~\ref{f:delomcl},\ref{f:deltsph}) or the high-K isomers (see
e.g. \cite{highk}).

In summary, at very low temperature an increasing external
magnetic field causes the magnetic moment of small
superconducting clusters ($R<5$ $nm$) to grow in a step-wise
manner. Each step reduces the pair correlations until they are
destroyed. However, with increasing temperature the steps are
washed out and substantial pair correlations re-appear for high
field strengths, where they are quenched at $T=0$. Nuclei that
built up angular momentum along a symmetry axis behave in an
analogous manner. The pair correlations are destroyed in a
step-wise manner by subsequent alignment of the angular momenta
of individual nucleonic orbitals with the symmetry axis. These
steps are washed out with increasing temperature and pair
correlations appear at values of the rotational frequency, where
they are quenched at $T=0$. This  phenomenon of
temperature-induced pairing reflects the strong fluctuations of
the order parameter in very small systems with a fixed particle
number.

Supported by the grants INTAS-93-151-EXT and DE-FG02-95ER40934.

\end{document}